\documentclass[times,twocolumn,final]{elsarticle}

\usepackage{medima}
\usepackage{framed,multirow}

\usepackage{amssymb}
\usepackage{latexsym}
\usepackage{subfigure}
\usepackage{url}
\usepackage[normalem]{ulem}

\setcitestyle{authoryear,open={(},close={)}}

\definecolor{newcolor}{rgb}{.8,.349,.1}

\journal{Medical Image Analysis}

\begin{document}

\verso{Andrea Mendizabal \textit{et~al.}}

\begin{frontmatter}

\title{Simulation of hyperelastic materials in real-time using Deep Learning}

\author[1,2]{Andrea Mendizabal \corref{cor1}}
\ead{andrea.mendizabal@inria.fr}
\cortext[cor1]{Corresponding author}

\author[3]{Pablo M\'arquez-Neila}
\author[1]{St\'ephane Cotin}

\address[1]{Inria, Strasbourg, France}
\address[2]{University of Strasbourg, ICube, Strasbourg, France}
\address[3]{ARTORG Center, University of Bern, Switzerland}


\begin{abstract}
The finite element method (FEM) is among the most commonly used numerical methods for solving engineering problems. Due to its computational cost, various ideas have been introduced to reduce computation times, such as domain decomposition, parallel computing, adaptive meshing, and model order reduction. 
In this paper we present U-Mesh: a data-driven method based on a U-Net architecture that approximates the non-linear relation between a contact force and the displacement field computed by a FEM algorithm. We show that deep learning, one of the latest machine learning methods based on artificial neural networks, can enhance computational mechanics through its ability to encode highly non-linear models in a compact form. Our method is applied to three benchmark examples: a cantilever beam, an L-shape and a liver model subject to moving punctual loads. 
A comparison between our method and proper orthogonal decomposition (POD) is done through the paper. The results show that U-Mesh can perform very fast simulations on various geometries and topologies, mesh resolutions and number of input forces with very small errors.
\end{abstract}

\begin{keyword}

\KWD \\
Real-time simulation \sep  
Deep neural networks \sep 
Physics-based simulation \sep
Finite element method \sep 
Hyperelasticity \sep 
Reduced order model
\end{keyword}

\end{frontmatter}

\section{Introduction}
\vspace{-0.3cm}
There are many applications in engineering where the deformation of non-linear  structures needs to be simulated in real time, or would benefit from being computed interactively. Some important examples can be found in the field of medicine, in order to develop training systems for learning surgical skills \citep{Ayache2006} or in the field of surgical navigation, where augmented reality combined with interactive simulations can bring significant improvements to clinical practice \citep{Haouchine2013}. Medical robotics, involving flexible robots or interactions with soft tissues, is another important area where real-time simulation of flexible structures is essential, in order here to have a better control of the robot.

While there are different numerical strategies for solving equations associated with elastic materials, we only consider the finite element method (FEM) in this article, for its accuracy and ability to simulate a large range of materials on potentially complex domains. However, obtaining real-time simulations with this method, in particular when considering non-linear materials, becomes a real challenge, in particular if this has to be done on consumers level hardware rather than a high-end parallel computer.

In order to speed up FEM simulations, several techniques have been proposed. We review here only some of the main ideas that were proposed. First, since solving the system of equations resulting from the FEM discretization is usually the bottleneck of the computation, many works have focused on linear solvers. 
Domain decomposition methods are based on the “divide $\&$ conquer” paradigm. Such methods consist in splitting the global problem domain into smaller independent sub-domains, making the approach suitable for parallel computing. 
This allows to build efficient preconditioners even though additional computation is required to synchronize the solution between neighboring sub-domains. 
Under the right conditions, in particular if the number of processors of the computer matches the number of sub-domains, a superlinear speedup can be obtained \citep{Haferssas2017}. This is, however, impossible to achieve when considering problems (even if relatively small) which need to be solved in real-time on consumers level hardware. This is mainly due to the limited number of cores (only 10 cores on the latest Intel i9 processor) and communication costs which are significant compared to the expected computation times (about 50 ms per time step for an interactive simulation).

Another option for speeding up simulation times is to lower the computational complexity of the problem through a reduction of the model's degrees of freedom. Depending on the problem, and acceptable loss of accuracy, it is possible to obtain speedups of several orders of magnitude. Proper Orthogonal Decomposition (POD) is one of the main model order reduction methods. 
POD techniques compute off-line the solution to several complete models and extract the modes that describe best the solution to the complete problem. Based on a priori knowledge of the solution, it encodes the high dimensional problem in a smaller subspace defined by a truncated basis of singular vectors. The dimension of such basis determines the ratio between accuracy of the method and computation times (the smaller the basis, the larger the error). In the context of real-time simulation of non-linear solids, several examples of POD have been proposed. \cite{Niroomandi2008} proposed a POD method to simulate the palpation of the cornea. Haptic feedback rates were achieved, but with a relative error of about 20\%. If more accurate solutions are needed, the number of modes used in the POD can be increased at the cost of higher computation times. Thus, to keep the method numerically efficient in the case of non linear materials,  hyperreduction can be used in order to further reduce the computation times while reducing the error \citep{hyper}. \cite{Goury2018} applied the hyperreduced POD to control and simulate soft robots with very good accuracy in $25~ms$ per time step.
However the POD may be in some cases insufficient to capture correctly the high degrees of non linearity that can be found for example in biological soft tissues as it relies on a linear combination of few basis vectors \citep{manifold}. To precisely account for  non-linearities it might be necessary to recompute the entire stiffness matrix which is burdensome and not always possible \citep{Niroomandi2017}.
 Another model order reduction algorithm is the proper generalized decomposition (PGD), which, contrarily to POD, builds a reduced-order approximation without relying on the knowledge of the solution of the complete problem. PGD assumes that the solution of a multiparametric problem can be expressed as a sum of separable functions that are constructed by successive enrichment by invoking the weak form of the considered problem.
 An approach based on PGD is proposed in \citep{Niroomandi2013} for the simulation of hyperelastic soft tissue deformation at high frequency. However, when the solution is non-separable, PGD offers no particular advantage over classical FEM techniques.

 A last class of worth mentioning solutions consists in using the Graphics Processing Unit (GPU) as a particular type of parallel machine. Although each core of the GPU is very limited in its computational performance, the very high number of cores available (several thousand) makes it possible to obtain significant speedups on computationally heavy problems. For instance, NiftySim \citep{Johnsen2015} is a GPU-based non-linear finite element toolkit for the simulation of soft tissue biomechanics where speedups of 300x are obtained. SOFA~\citep{sofa} is an Open-source Framework focused on real-time simulation of complex interactions with deformable structures, which provides GPU-compatible FEM codes \citep{Comas2008}. Based on a Total Lagrangian Explicit Dynamics algorithm from \citep{Miller2007}, a speedup of more than 50x is reported.


Recently, machine learning started to revolutionize several fields (vision, language processing, image recognition, genomics) due to the continuously increasing amount of data available and the development of new algorithms and powerful GPUs. Deep learning, a class of machine learning methods based on learning data representations, as opposed to task-specific algorithms, has demonstrated strong abilities at extracting high-level representations of complex processes. With sufficient ground truth data, machine learning algorithms can map the input of a function to its output without any mathematical formulation of the problem, thus actuating like a black box. Since FEM can provide as much noise-free data as required, it seems interesting to train learning algorithms with such virtually generated data \citep{Lorente2005,Luo2018,Roewer-despres2018,Tonutti,Fetene,Runge}. For example \cite{Lorente2005} proposed a machine learning approach for modeling the mechanical behavior of the liver during breathing in real-time. They trained several regression models using external displacement and elasticity parameters as input, and the FEM-based nodal displacements as output. Although they reached good accuracy their method is restricted to small displacements. \cite{Roewer-despres2018} proposed a preliminary work using a deep-autoencoder to approximate the deformations of a non-linear muscle actuated object. They showed that their method produces lower reconstruction errors than the equivalently sized PCA model. However the computational gain of their method is not clear and it was limited to a very simple model and coarse mesh. \cite{Tonutti} treated a simple problem using two different networks, one to predict the magnitude of the displacement and the other one to predict its direction. However, it seems that these two networks require a specific training for each node of interest, which could be prohibitive for large meshes. Moreover their model is limited to small deformations on relatively simple shapes, with restricted input forces. For instance, the considered displacements never exceed 5 mm (for an organ of size 20 cm) and only 11 nodes of the mesh are excited. On the contrary our approach can handle complex and complete volume deformations of arbitrary shapes with one single network for any force application point. All the cited references propose to train neural networks with FEM generated data for various purposes. However, none of them is justifying the choice of the network architecture used to do this. In this paper, we explore connections of the chosen architecture with model order reduction techniques in order to support interpretability and explainability to go beyond the black-box usage of neural networks.

The objective of our work, whose preliminary results are presented in this article, is to go beyond the state of the art on machine learning applied to computational mechanics. In particular we show that we can predict, in real-time, the shape of a non-linear elastic structure with a very good accuracy using a deep network inspired by model order reduction methods. Our solution relies on a U-Net architecture trained on FEM-generated data sets; both aspects are presented in section \ref{sec:method} while results and their comparison against a model order reduction algorithm are presented in sections \ref{sec:results} and \ref{sec:discussion}.

\section{Method}
\label{sec:method}
As mentioned in the introduction, an important area of applications for real-time simulation of non-linear material is in the field of computer-aided surgery. In this context, accuracy is also very important but often left aside for the sake of rapidity. Achieving a better trade-off between accuracy and computation time is therefore mandatory to tackle more ambitious problems. This requirement for both accuracy and very fast computation can find applications in other areas of mechanical engineering, such as training of complex industrial processes \citep{aircraft} or virtual prototyping \citep{barbic} just to name a few.

In this paper, we propose a method that does not need such a compromise. It allows for extremely fast and accurate simulations by using an artificial neural network that partially encodes the stress-strain relation in a low-dimensional space. Such a network can learn the desired biomechanical model, and predict deformations at haptic feedback rates with very good accuracy. This section is divided in three main segments. First the problem that we aim to solve is presented, with the corresponding modeling and discretization choices. Then, the selected network architecture is detailed, followed by our strategy to encode the stress-strain relationship and boundary conditions into the network and the data set generation used to train the network.  

\subsection{Mechanical formulation of the problem and offline numerical resolution}

Without lack of generality, we consider the boundary value problem of computing the deformation of a hyperelastic material under both Dirichlet and Neumann boundary conditions. The solid occupies a volume $\Omega$ whose boundary is $\Gamma$. We assume the Dirichlet conditions on $\Gamma_D$ known a priori, while Neumann boundary conditions on $\Gamma_N$ can change at any time step. We consider a Lagrangian description of the deformation whose material coordinates are given by the vector $\textbf{X}$. The deformed state of each point of the solid is given by
    \begin{equation}
    \textbf{x}=\textbf{X}+\textbf{u}
    \end{equation}
where $\textbf{u}$ is the displacement field.
We propose to describe the linear relation between the stress and the strain using a Saint-Venant-Kirchhoff constitutive model. The Green-Lagrange strain tensor $\textbf{E} \in \mathbf{R}^{3\times 3}$ is computed as a non-linear (quadratic) function of the deformation gradient $\textbf{F} = \textbf{I} + \nabla_{\textbf{X}} \textbf{u}$:
\begin{equation}
        \textbf{E}= \frac{1}{2}(\textbf{F}^T\textbf{F} - \textbf{I})
        \label{eq:strain}
\end{equation}
where $\textbf{I} \in \mathbf{R}^{3\times 3}$ is the identity matrix. The strain-energy density function $\textbf{W}$ is obtained according to the following equation:
    
\begin{equation}
\textbf{W}(\textbf{E})=\frac{\lambda}{2}[tr(\textbf{E})]^2 + \mu tr(\textbf{E}^2)
\end{equation}
where $\lambda$ and $\mu$ are the Lame's constants derived from the Young's modulus $Y$ and the Poisson's ratio $\nu$. 

Then the second Piola-Kirchhoff stress $\textbf{S}$ reads as follows:
	\begin{equation}
	\textbf{S}=\frac{\partial \textbf{W(\textbf{E})}}{\partial \textbf{E}}=\textbf{C}:\textbf{E}
	\end{equation}
where $\textbf{C}$ is the fourth-order constitutive tensor. $\textbf{S}$ is related to the first Piola-Kirchhoff stress tensor $\textbf{P}$ by $\textbf{P}=\textbf{F}\textbf{S}$.

Ignoring time-dependant terms, the boundary value problem is then given in material coordinates by :

\begin{equation}
\left\{
\begin{array}{ll}
    \nabla (\textbf{FS}) = \textbf{b}~in~\Omega \\
    \textbf{u}(\textbf{X}) = 0~on~ \Gamma_D\\
    (\textbf{FS}) \textbf{n} = \textbf{t} ~on ~\Gamma_N 
\end{array}
\right.
\label{eq:bvp}
\end{equation}
where $\textbf{b}$ is the external body force, $\textbf{n}$ is the unit normal to $\Gamma_N$ and $\textbf{t}$ is the traction applied to the boundary. The weak form of (\ref{eq:bvp}), obtained from the principle of virtual work, brings forward the boundary term and reads as:
\begin{equation}
    \int_\Omega (\textbf{FS}) : \delta \textbf{E}~ d\Omega = \int_\Omega \textbf{b} \mathbf{\eta} ~d\Omega + \int_{\Gamma_N} \textbf{t} \mathbf{\eta} ~d\Gamma
\label{eq:weak}    
\end{equation}
where $\delta \textbf{E} = \frac{1}{2}(\textbf{F}^T \nabla \mathbf{\eta} + \nabla^T \mathbf{\eta} \textbf{F})$ is the variation of the strain, and $\mathbf{\eta} = \{\mathbf{\eta}~\in H^1(\Omega) ~|~ \mathbf{\eta} = 0  ~on~  \Gamma_D\}$ is any vector-valued test function ($H^1(\Omega)$ being a Hilbert space). The left side of equation (\ref{eq:weak}) denotes the internal virtual work, and the right side, the virtual work from the applied external load.

\paragraph{Finite element simulation}\label{par:fem}
    Equation (\ref{eq:weak}) is solved using a finite element method. The domain was discretized in hexahedral (H8) elements, providing a set of unknown displacements at the element nodes. This choice is not only motivated by the good convergence and stability of such elements: hexahedral elements are also required for our convolutional neural network (see section \ref{sec:unet}).
    
    Due to the non-linearity of equation (\ref{eq:strain}), we need to solve a non-linear system of equations to approximate the unknown displacement.
    Using an iterative Newton-Raphson method, from an initial displacement $\textbf{u}^0$, we 
    try to find a correction $\mathbf{\delta}_u^n$ after $n$ iterations that balances the linearized set of equations:
    \begin{equation}
    \label{eq:discrete}
    \dot{\textbf{K}}^{n-1} \mathbf{\delta}_u^n = \textbf{r}(\textbf{u}^0 +\mathbf{\delta}_u^{n-1}) + \textbf{b}
    \end{equation}
    where $\dot{\textbf{K}}$ is the tangent stiffness matrix and $\textbf{r}$ is the internal elastic force vector. Here, at each iteration, both the matrix $\dot{\textbf{K}}$ and the vector $\textbf{r}$ need to be computed, and the linear system needs to be solved. Since the convergence of the Newton-Raphson method is only valid for a displacement $\textbf{u}^0$ near the solution, large external loads must be applied by small increments and can require a large number of iterations to converge. This is an important characteristic in order to evaluate the performance of the proposed neural network on a non-linear model.       
    Depending on the mesh resolution, solving this problem can be extremely long even for very sophisticated and optimized codes. Dimensionality reduction techniques have shown real benefits in speeding-up FEM simulations. Among them, POD is a very popular one since it leads to very realistic real-time simulations. Although authors in \citep{Niroomandi2008} have shown good results using this method for large deformation, POD is better suited for linear or weakly non-linear processes. We propose a non-linear dimensionality reduction technique using a neural network to learn the correspondence from contact forces to volumetric deformations of a given mesh.

\subsection{Deep neural network for online prediction of the displacement field}
\label{sec:unet}
        Formally, our network~$h$ is a parameterized function that accepts a $3\times{}n_x\times{}n_y\times{}n_z$ tensor~$\mathbf{f}$ as input and produces a tensor~$\mathbf{u}$ of the same size as output. The domain $\Omega$ is sampled by a 3-dimensional grid of resolution $n_x\times{}n_y\times{}n_z$. Practically speaking, the nodes of this grid match the nodes of the FEM mesh, although this is not  required and an interpolation could be used instead. The tensor $\mathbf{f}$ represents the contact forces applied to the mesh, and the tensor~$\mathbf{u}$ contains the corresponding displacement of the mesh. In particular, each vector~$\mathbf{f}[:, i, j, k]$ represents the force vector~$(f_x, f_y, f_z)$ applied over the node $(i, j, k)$ of the grid. Similarly, the vector~$\mathbf{u}[:, i, j, k]$ represents the displacement~$(u_x, u_y, u_z)$ of the node $(i, j, k)$.

        Our problem consists of finding the function~$h$ that produces the best estimations of the displacement field given some contact forces (traction). This is performed by minimizing the expected error over the distribution~$\mathcal{D}$ of pairs~$(\mathbf{f}, \mathbf{u})$,
        
        \begin{equation}
          \label{eq:loss}
          \min_{\theta} \mathbb{E}_{(\mathbf{f}, \mathbf{u})\sim\mathcal{D}}\left[ \|h(\mathbf{f}) - \mathbf{u}\|^2_2 \right],
        \end{equation}
        where $\theta$ is the set of parameters of the network~$h$.
        In practice, the expectation of Eq.~\ref{eq:loss} is approximated by Monte-Carlo sampling with a training set~$\{(\mathbf{f}_n, \mathbf{u}_n)\}_{n=1}^N$ of $N$~samples:
        \begin{equation}
          \label{eq:loss2}
          \min_{\theta} \frac{1}{N}\sum_{n=1}^N \|h(\mathbf{f}_n) - \mathbf{u}_n\|^2_2.
        \end{equation}
        
        We build our training set by randomly applying forces on the mesh and running FEM simulations to produce corresponding displacements. 
       
\begin{figure}
    \centering
\includegraphics[width=0.5\textwidth]{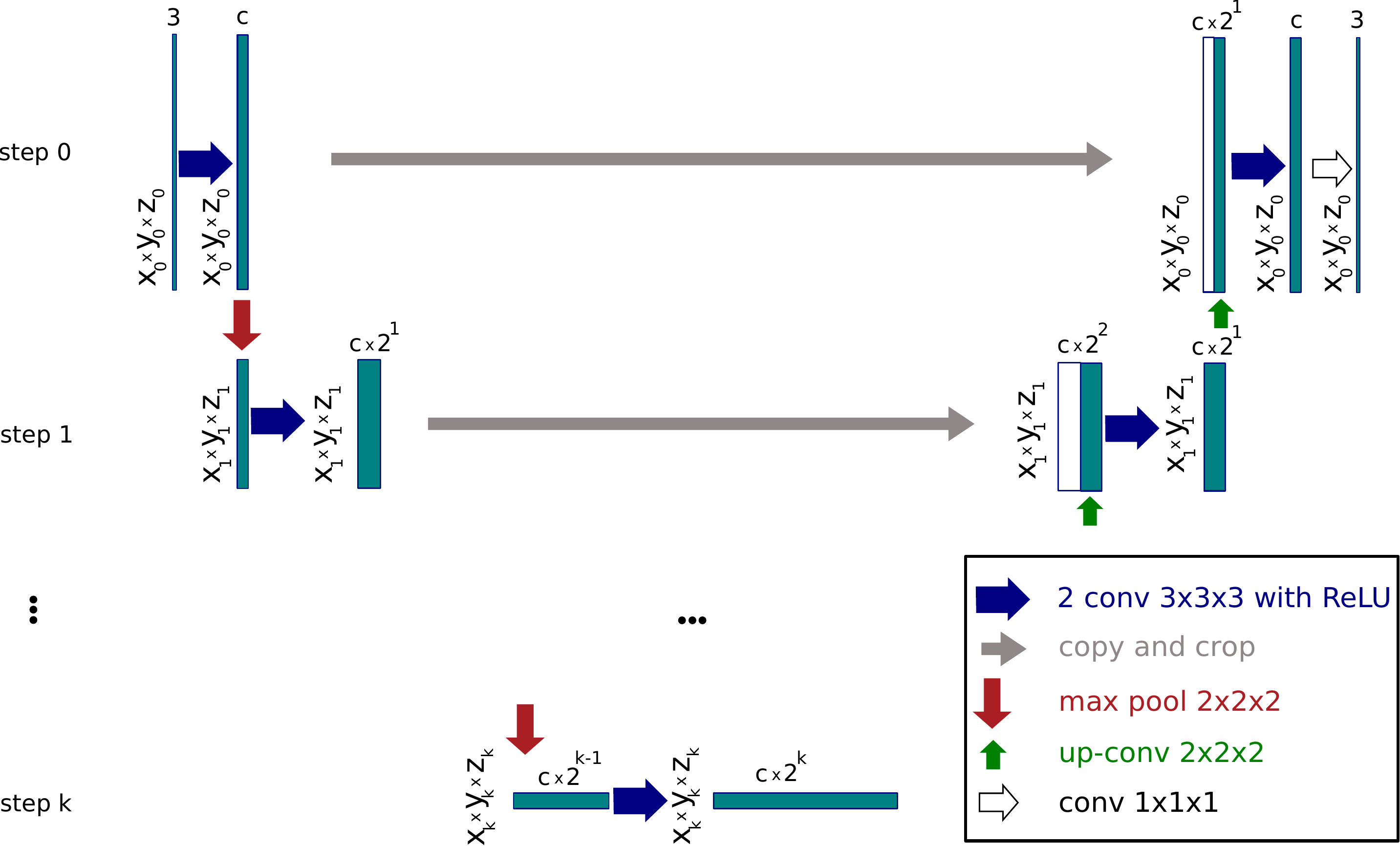}
    \caption{General network architecture for an object with a resolution of $x\times y \times z$ nodes, $c$ channels in the first layer and $k$ steps. Note that $x_0\times y_0 \times z_0$ is the padded grid.}
    \label{fig:unetgraph}
    \end{figure}

        Let us characterize the architecture chosen for our network~$h$. We propose to use the U-Net~\citep{Ronneberger2015}, a modified fully convolutional network initially built for precise medical image segmentation. As depicted in Fig.~\ref{fig:unetgraph}, the network is similar to an auto-encoder, with an encoding path to transform the input space into a low-dimensional representation, and a decoding path to expand it back to the original size. Additional skip connections transfer detailed information along matching levels from the encoding path to the decoding path.

\begin{figure}
    \centering
    \includegraphics[width=0.5\textwidth]{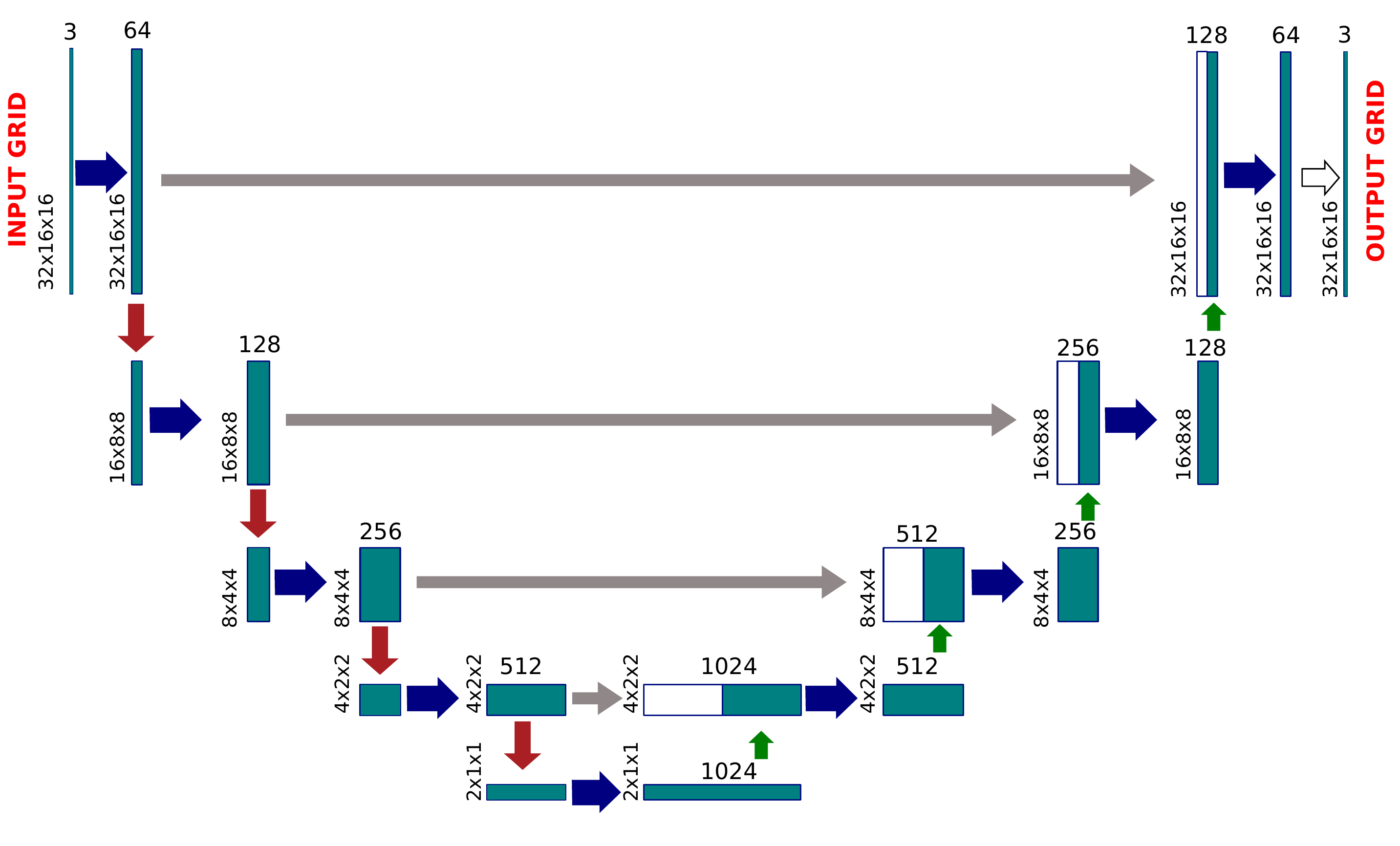}
    \caption{Network architecture for a beam with $28 \times 12 \times 12$ nodes, padded to $32 \times 16 \times 16$, 64 channels in the first layer and 4 steps (see Fig.~\ref{fig:unetgraph} for notations).}
    \label{fig:unetgraph_for_beam}
    \end{figure}    
    
    The encoding path consists of $k$~sequences of two padded $3\times{}3\times{}3$ convolutions ($k=4$ in~\citep{Ronneberger2015}) and a $2\times{}2\times{}2$~max pooling operation (see Fig.~\ref{fig:unetgraph}).
    Intuitively, each 3D convolution filter learns to isolate the different characteristics of the displacement field (orientation, direction, amplitude).
At each step, each feature map doubles the number of channels and halves the spatial dimensions. We assume that the number of channels is directly related to the amount of detectable variations in the displacement field. In the bottom part there are two extra $3\times{}3\times{}3$~convolutional layers leading to a ($c\times 2^{k}$)-dimensional array. This feature space is similar to the Galerkin projection of the equations of motion onto the reduced space in POD, where the order of the singular vector truncation is equivalent to the number of neurons in the latent space. A difference however remains with the presence of convolutional operations at each layer.
In a symmetric manner, the decoding path consists of $k$~sequences of an upsampling $2\times{}2\times{}2$~transposed convolutions followed by two padded $3\times{}3\times{}3$~convolutions. The features from the encoding path at the same stage are cropped and concatenated to the upsampled feature maps. At each step of the decoding path, each feature map halves the number of channels and doubles the spatial dimensions. There is a final $1\times{}1\times{}1$~convolutional layer to transform the last feature map to the desired number of channels of the output (3 channels in our case). 
    
    The number of steps~$k$ and the number of channels~$c$ control the accuracy of the prediction just like the number of singular vectors in POD. Higher values of $k$ and~$c$ lead to a more complex network suitable for difficult meshes at the expenses of longer computational times for both training and prediction, and higher requirements of training data. We tested several values for~$k$ and $c$ in order to select the most appropriate values for our experiments depending on the desired accuracy and the eventual time restrictions. 

\subsection{Data set generation for U-Net training}

         In order to train such a network, we build a data set of pairs $(\mathbf{f},\mathbf{u})$ obtained with the previously explained FEM.
     Once we are given a 3D mesh with its corresponding constitutive law, material properties and boundary conditions, we perform multiple simulations by applying random forces to nodes of the object. After each simulation, the pair of applied forces and obtained deformation is stored as an element of the data set. To speed up the generation of the training and testing data sets, the linearized system of equation (\ref{eq:discrete}) is solved using an iterative preconditioned conjugate gradient method~\citep{Shewchuk}. 
         
         The variability of the data relies on the force magnitude, its direction and its application point. In this work, a force is applied on a local surface area $\Gamma_N$ whose location varies such that the boundary of the computation domain is completely covered. The force direction is uniformly sampled on the unit sphere and the force magnitude is a uniform random value between 0 and~1. At each sample of the data set, one force is applied on a small region $\Gamma_N$. There are $\Lambda$ samples for each $\Gamma_N$, meaning that $\Lambda$ different forces are applied per region. We can consider one force per sample or several forces applied simultaneously at different locations. 
         
          \paragraph{Training}
        The network is trained minimizing Eq.~\ref{eq:loss2} with training data generated as explained above. The minimization is performed using the Adam optimizer~\citep{KingmaB14}, a stochastic gradient descent procedure with parameter-wise adjusted learning rates.

\section{Results}
\label{sec:results}

In this section, we perform a model selection over the space of hyperparameters~$k$ and~$c$ in order to find the combination that leads to the best results in a cantilever beam. Then we apply our method, with the selected hyperparameters, to three benchmark examples: a cantilever beam, an L-shaped object and a liver shape under point loads. 
All our experiments are performed in a GeForce 1080 Ti using a batch size of $4$ and $100,000$ iterations for training. We use a PyTorch implementation of the U-Net. We recall that the batch size is the number of samples that are given to the network at each iteration of the minimization process.


 \subsection{Validation metrics}

To assess the efficiency of our method, we perform a statistical analysis of the \textit{mean norm error} $e$ over a testing data set~$\{(\mathbf{f}_m, \mathbf{u}_m)\}_{m=1}^M$, which is built similarly as the training data set. Note that the training data set and the testing data set are disjoint. Let $\mathbf{u}_m$ be the ground truth displacement tensor for sample $m$ generated using the FEM described in section \ref{par:fem} and $h(\mathbf{f}_m)$ the U-Mesh prediction.
The mean norm error between $\mathbf{u}_m$ and $h(\mathbf{f}_m)$ for sample $m$ reads as:
\begin{equation}
    e(\mathbf{u}_m,h(\mathbf{f}_m)) = \frac{1}{n}\sum_{i=1}^n |\mathbf{u}_m^i-h(\mathbf{f}_m)^i|.
\end{equation}
where $n$ is the number of degrees of freedom of the mesh. We compute the average $\overline{e}$ and standard deviation $\sigma(e)$ of such norm over the testing data set:
\begin{equation}
    \overline{e} = \frac{1}{M}\sum_{m=1}^M e(h(\mathbf{f}_m), \mathbf{u}_m),
\end{equation}
and
 \begin{equation}
    \sigma(e) = \sqrt{\frac{1}{M-1}\sum_{m=1}^M \left[e(h(\mathbf{f}_m), \mathbf{u}_m)-\overline{e}~\right]^2}.
\end{equation}

\subsection{U-Mesh applied to a cantilever beam}
 
\begin{figure}
                \centering
                \includegraphics[width=0.32\textwidth]{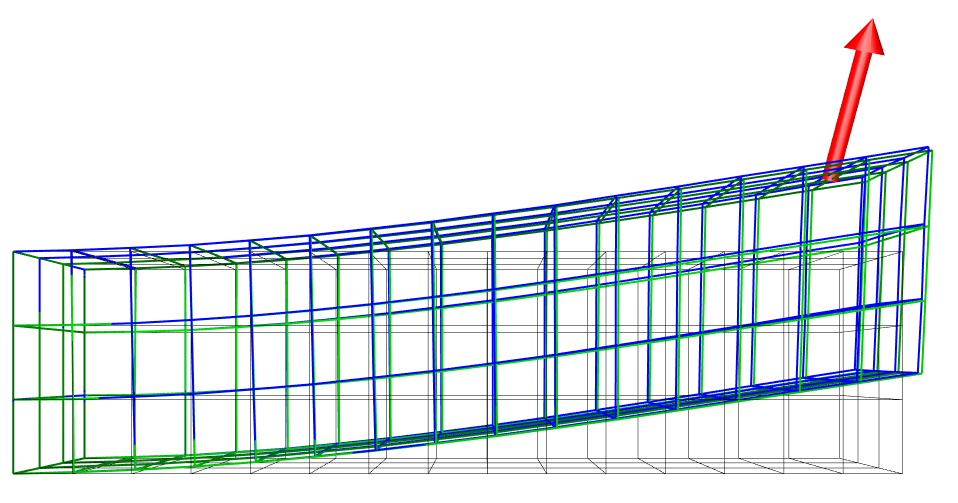}
                \includegraphics[width=0.15\textwidth]{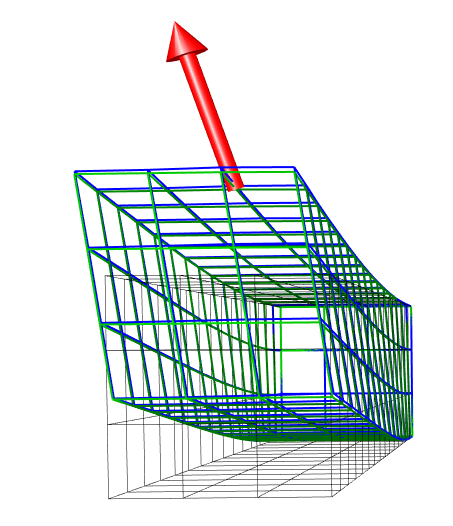}
            \caption{Front and side views of the maximal error sample obtained with a network having 3 steps and 128 channels. U-Mesh output is in blue and the reference is in green. The red arrow represents the force applied. Note that the difference between the two meshes is very small. The relative $l2$ norm at the tip of the beam is of $5.1\%$ and the deformation amplitude is $0.45\,m$}
                \label{he}
            \end{figure}

\begin{figure}
                \centering
                 \includegraphics[width=0.5\textwidth]{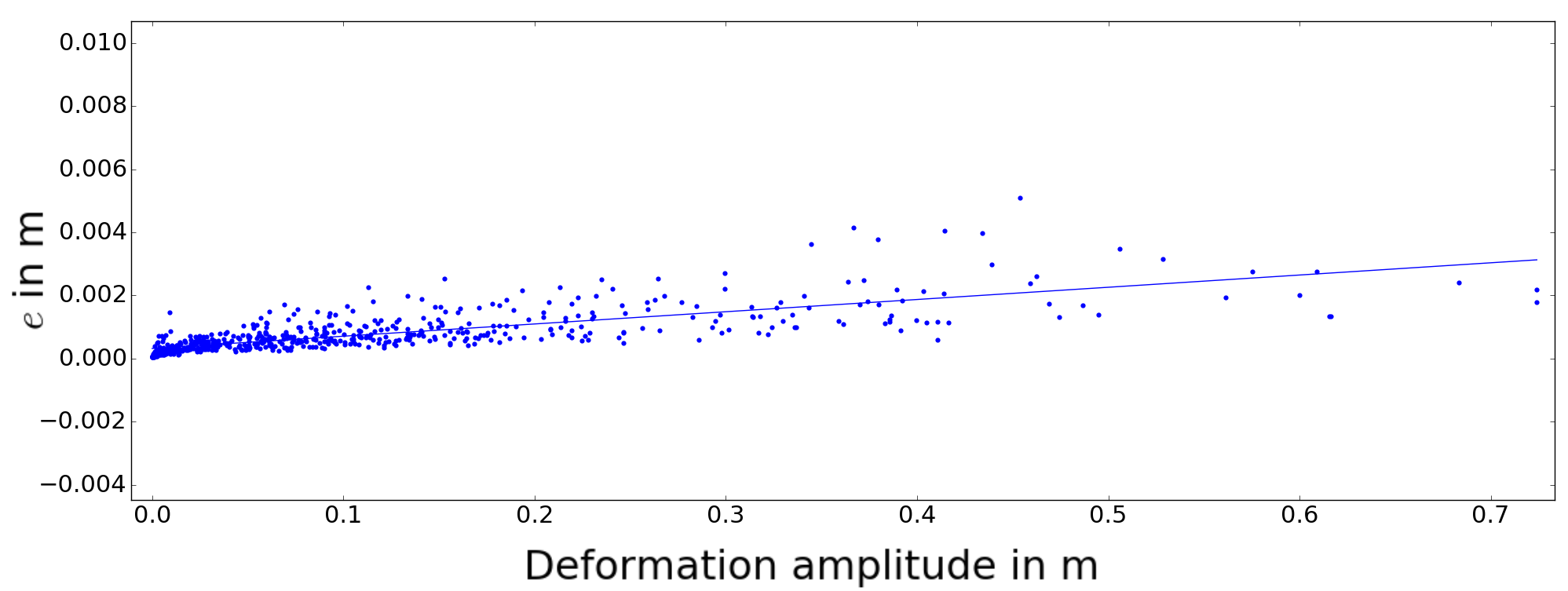}
            \caption{The point cloud represents the $e$ for some randomly selected samples of the testing data set. The regression line of equation $y=0.00352\times x$ shows the low sensitivity of the U-Mesh to the deformation amplitude.}
                \label{fine_errors}
            \end{figure}

We consider a deformable beam of size $4  \times 1  \times 1\,m^3$ subjected to fixed boundary on one end. The beam follows the Saint-Venant-Kirchhoff behavior described in section \ref{sec:method} with a Young's modulus $Y$ of $500$ Pa and a Poisson's ratio of $0.4$, and is discretized with 135 H8 elements. 
To generate the data set, 100 forces are applied on each of the 64 nodes of the upper face (that is $\Lambda = 100$), up to a total of $6,400$ samples. $80 \%$ of the samples are used for training ($N = 5,120$) and the remaining $20 \%$ are kept for testing ($M=1,280$). 
Using this data set, we select the best machine learning model by training several U-Net architectures with different combinations of hyperparemeters $k$ and~$c$.
In Table~\ref{beam256_1f} are reported the training and prediction times as well as $\overline{e}$ and $\sigma(e)$. 
As seen in this table, the higher the feature space size (FSS), the lower the errors. Prediction time is proportional to the depth of the network.
Choosing the best model is a tradeoff between network performance and speed, and we will show that the selected hyperparameters lead to good results also on different problems.
Choosing $k=3$ and $c=128$ seems to be a good compromise for our needs.
The selected parameters appear in bold text in Table~\ref{beam256_1f}. For the selected set of parameters, $e$ over the data set is equal to $0.0007 \pm 0.0006\,m$ for a maximal deformation of $0.724\,m$. In Fig.~\ref{he}, we show the sample with maximal error.
We perform a sensitivity analysis of the method to the amplitude of deformation. The results are plotted in Fig.~\ref{fine_errors}. We perform a least squares line fit to find the relation between the maximal deformation and the mean norm error $e$. We can observe a very small sensitivity of the error $e$ with respect to the deformation amplitude, and a very small error in the estimation of the displacement field in general. This shows an important characteristic of our method, in addition to its very limited computational cost.

             \begin{table}
            \begin{center}
            \footnotesize
              \begin{tabular}{ |c|c|c||c|c||c|c| } 
               \hline
               c & k & FSS  & $\overline{e}$ & $\sigma$($e$)  & pred\_t  & train\_t \\ 
                 &  & & in m  & in m & in ms &  in min\\ 
                \hline
            64&2&256 &0.0028&0.0008&2&24\\ 
             16&4&256 &0.0012&0.0009&3.2&40\\
              32&4&512 &0.0009&0.0008&3.2&40\\
             64&3&512 &0.0007&0.0007&2.5&80\\
             64&4&1024 &0.0007&0.0007&3.3&95\\
             \textbf{128}&\textbf{3}&\textbf{1024} &\textbf{0.0007}&\textbf{0.0006}&\textbf{2.5}&\textbf{35}\\
            64&5&2048 &0.0006&0.0005&4&426\\
             128&4&2048 &0.0006&0.0005&3.45&320\\
               \hline
              \end{tabular}
            \caption{Error measures for a beam having 135 H8 elements. Rows are sorted in decreasing $\overline{e}$ . The selected architecture appears in bold.}
            \label{beam256_1f}
            \end{center}
            \end{table}

\begin{figure}
                \centering
                \includegraphics[width=0.33\textwidth]{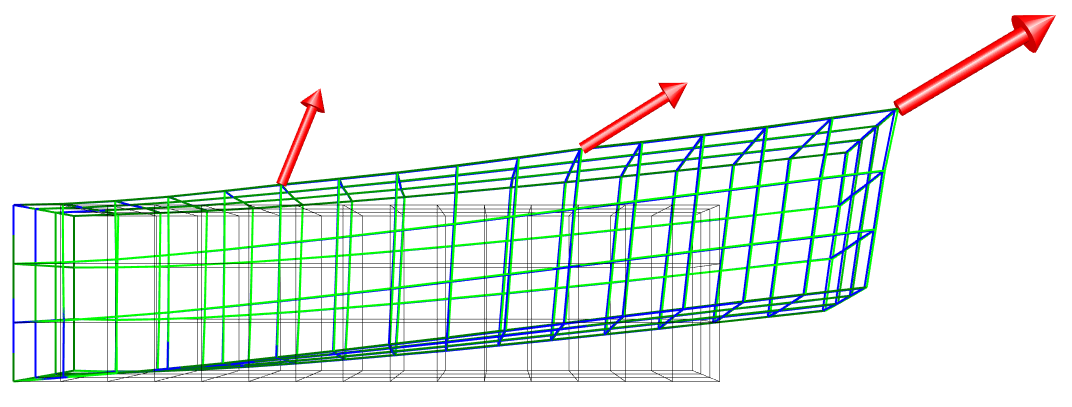}
                \includegraphics[width=0.14\textwidth]{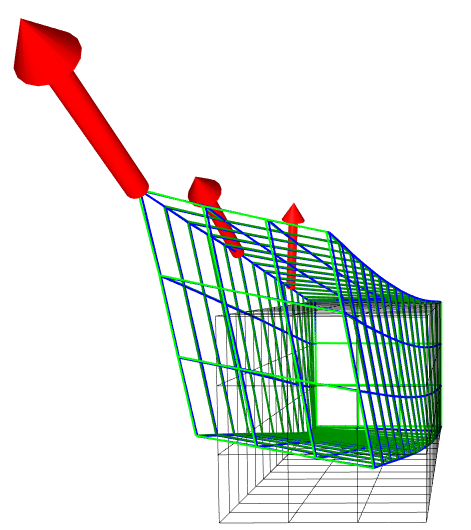}
            
                \caption{Largest deformation for 3 simultaneous force applications (side and front view for the same sample). U-Mesh output is in blue, the reference is in green and the rest shape is in grey.}
                \label{3flarge}
            \end{figure}

In Table \ref{beam256_3f} are shown the prediction errors for 3 simultaneous input forces and their corresponding training and prediction times. In this scenario, the Young's modulus is set to $400~ Pa$ and three forces are applied simultaneously on three different regions of the upper face of the beam. In order to avoid mechanical coupling, such regions must be far enough from  each other. There are $12,360$ possible combinations of nodes fulfilling the above stated condition. Only one force ($\Lambda = 1$) is applied per viable combination up to a total of $12,360$ samples ($N=9,888$ samples for training and $M=2,472$ samples for testing). We can see that the errors and the time needed for the prediction are comparable to that needed for only one force application. It is important to note that there are 2x more samples in this data set (than in previous ones) due to the large number of possible combinations. This explains the particularly low error in this case. The sample with maximal deformation ($1.0035~ m$) is shown in Fig.~\ref{3flarge}. At the tip of the beam, the relative $l2$ norm is equal to $1.5\%$.
    
            \begin{table}
            \begin{center}
              \begin{tabular}{ |c|c|c||c|c||c|c| } 
               \hline
               c & k & FSS  & $\overline{e}$ & $\sigma$($e$)  & pred\_t  & train\_t \\ 
                 &  & & in m  & in m & in ms &  in min\\ 
             \hline
            64&4&1024 &0.0007&0.0003&3.5&100\\
               \hline
              \end{tabular}
            \caption{Error measures for a beam having 135 H8 elements and three simultaneous force application.}
            \label{beam256_3f}
            \end{center}
            \end{table}

So far we have seen that U-Mesh is able to predict the displacement field due to one or several forces with high accuracy and in an extremely short amount of time. Nevertheless, FEM codes are also able to compute the solutions on such meshes in limited computation times, even for hyperelastic models. Hence, in order to put ahead the real contribution of our work, we test our method on a computationally expensive problem.

\begin{figure}[h]
                \centering
        \includegraphics[width=0.32\textwidth]{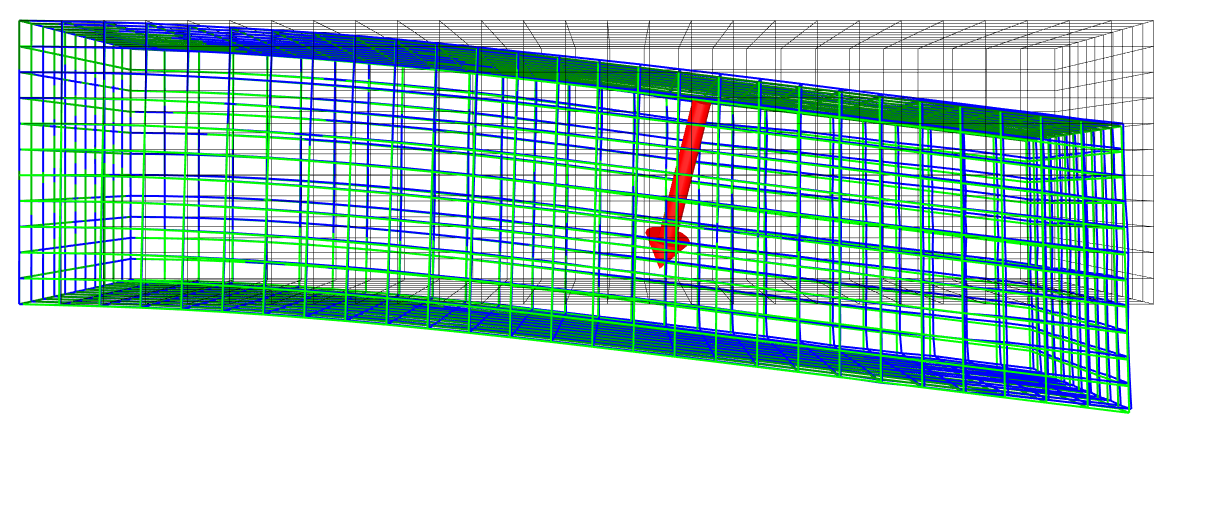}
        \includegraphics[width=0.15\textwidth]{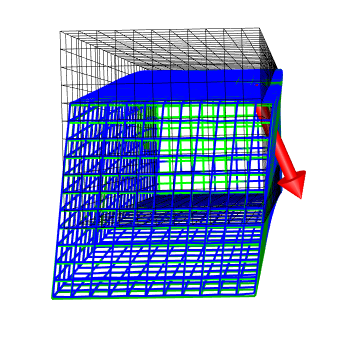}
    \caption{Beam mesh with 3267 H8 elements deformed with U-Mesh, front and side views. U-Mesh output is in blue, the reference is in green and the rest shape is in grey. The computation time is equal to $2.9\,ms$ for this sample. The relative $l2$ norm at the tip of the beam is $1.6\%$ and the deformation amplitude is $0.4\,m$ }
                \label{fig:fine}
            \end{figure}

We consider the same beam as previously but this time discretized in $3,267$ H8 elements (see Fig.~\ref{fig:fine}). There are 336 nodes on the upper face and $\Lambda$ is set to 100. Overall the data set has $33,600$ samples ($N=26,880$ samples for training and $M=6,720$ samples for testing). The U-Net is trained with the previously selected parameters, that is, 128 channels for the first layer and 3 steps. The metrics computed over the testing data set are shown in Table \ref{fine_beam_results}. The most interesting result is the low prediction time (only $3\,ms$). A very optimized version of the Saint-Venant-Kirchhoff FEM using a Pardiso solver \citep{pardiso} that is among the most  efficient solvers available, takes more than $300\,ms$ to solve such simulation. The speedup obtained with U-Mesh is of 100x. All the samples of the testing data set have an error bellow $0.0265\,m$ and an average error of $0.0019\,m$ for a maximal deformation of $1.011\,m$. 

             \begin{table}
            \begin{center}
              \begin{tabular}{ |c|c|c||c|c||c|c| } 
               \hline
               c & k & FSS  & $\overline{e}$ & $\sigma$($e$)   &pred\_t  & train\_t \\ 
                 &  & & in m  & in m & in ms &  in min\\ 
                \hline
             128&3&1024 &0.0019&0.0018&3&210\\
               \hline
              \end{tabular}
            \caption{Error measures for a beam having 3267 H8 elements and one simultaneous force application. }
            \label{fine_beam_results}
            \end{center}
            \end{table}

            \begin{figure}
                \centering
                \includegraphics[width=0.5\textwidth]{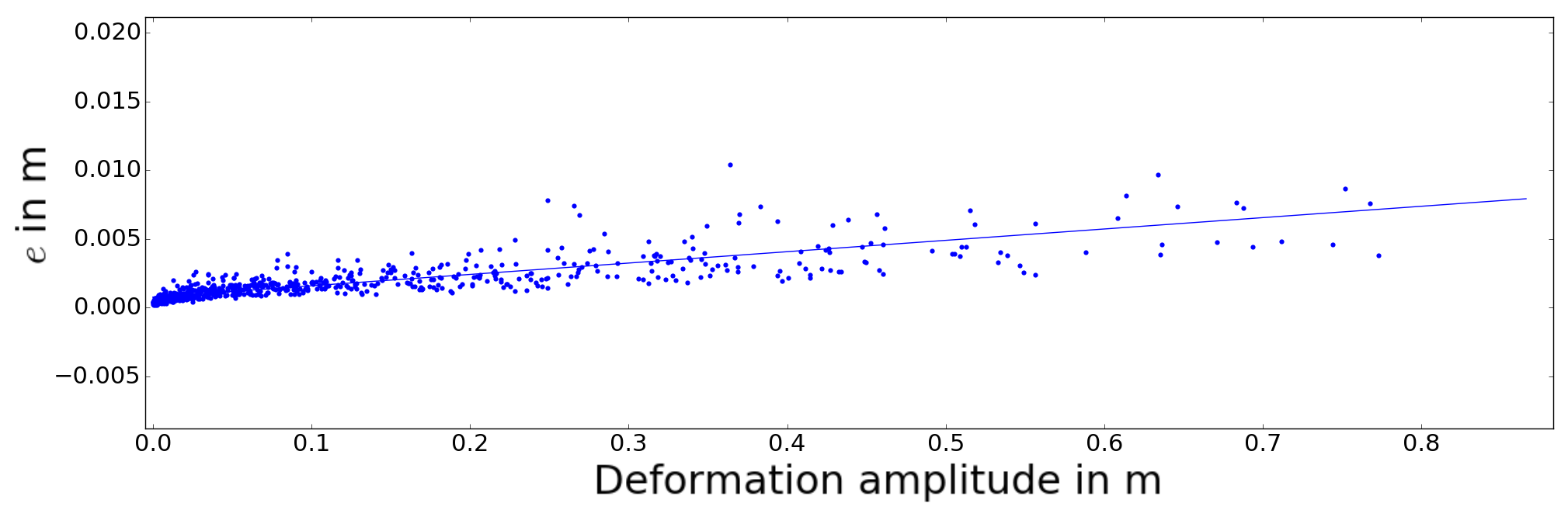}
        \caption{Sensitivity of $e$ to the deformation amplitude for a $3,267$ H8-elements beam. The point cloud represents the values of $e$ for randomly selected samples of the testing data set. The regression line of equation $y=0.0083\times x$ shows the low sensitivity of the U-Mesh to the deformation range. The average computation time is $3\,ms$.}
                \label{fig:errors_fine}
            \end{figure}

In the following paragraphs we will show that U-Mesh generalizes well on other geometries in particular, we will see the performance of U-Mesh applied to an L-shaped object and to a liver immersed in a regular grid.

\subsection{U-Mesh applied to an L-shaped object }
\begin{figure}
  \includegraphics[width=0.17\textwidth]{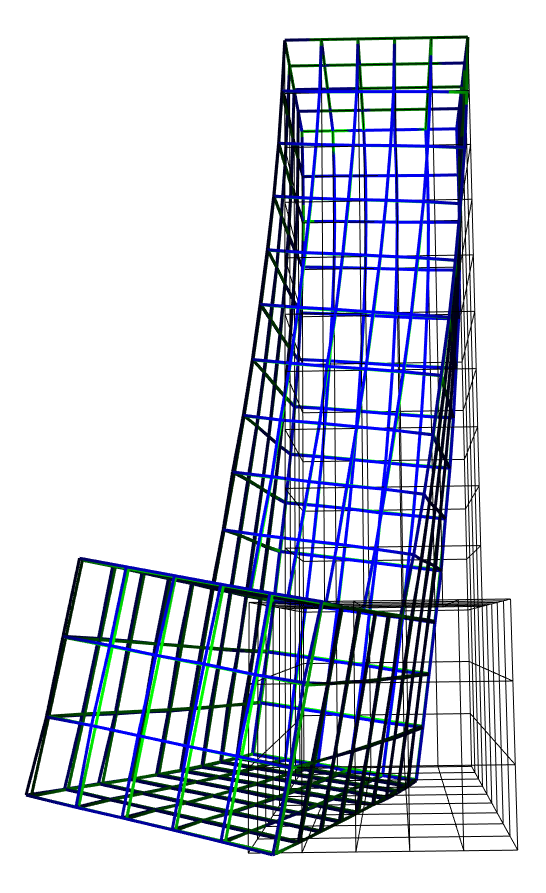}\includegraphics[width=0.215\textwidth]{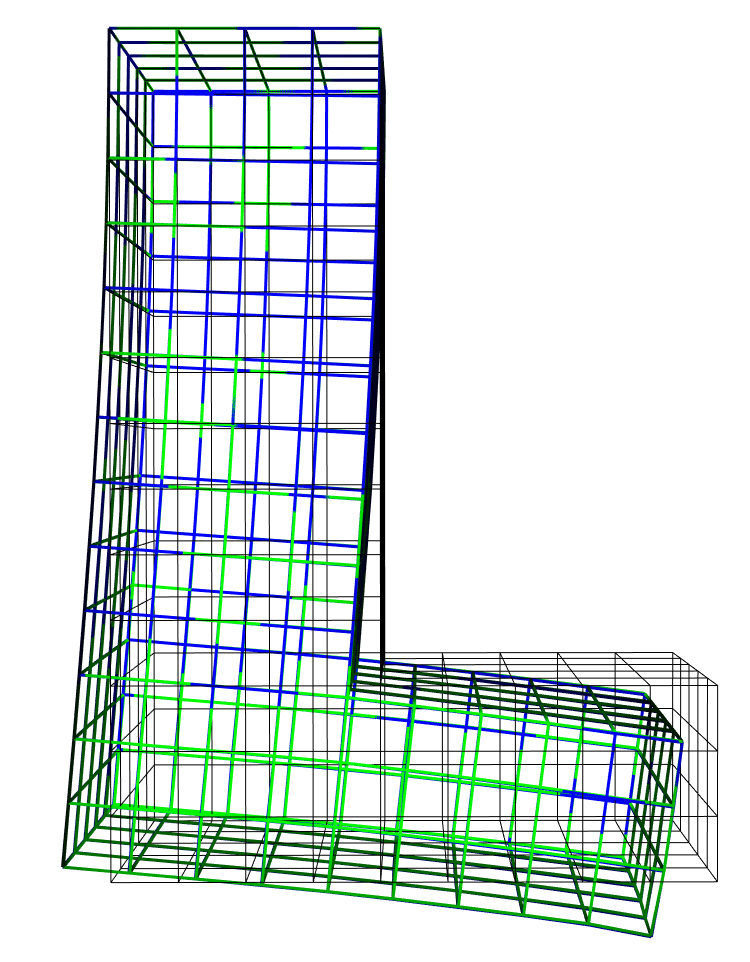}\includegraphics[width=0.14\textwidth]{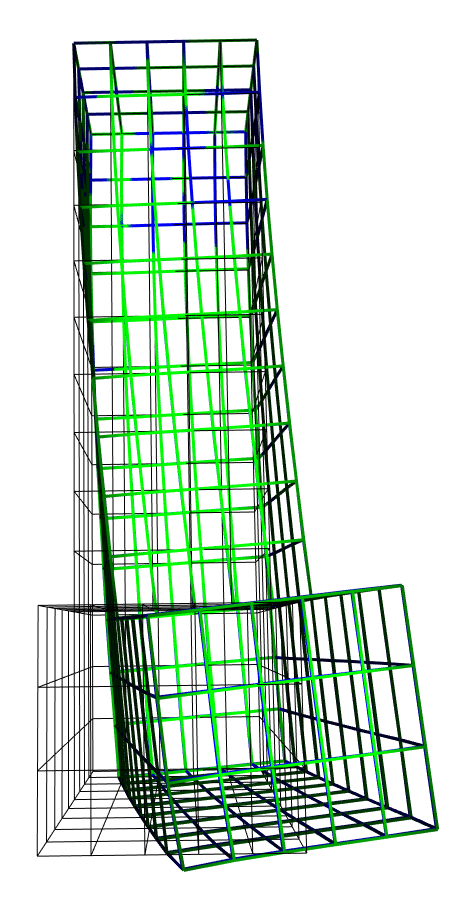}
  \caption{Samples of deformed L-shapes. In green is shown the reference solution, in blue the output of U-Mesh and in grey the rest shape.}
  \label{fig:L_bounded_40}
\end{figure}

\begin{figure}[h]
                \centering
                \includegraphics[width=0.5\textwidth]{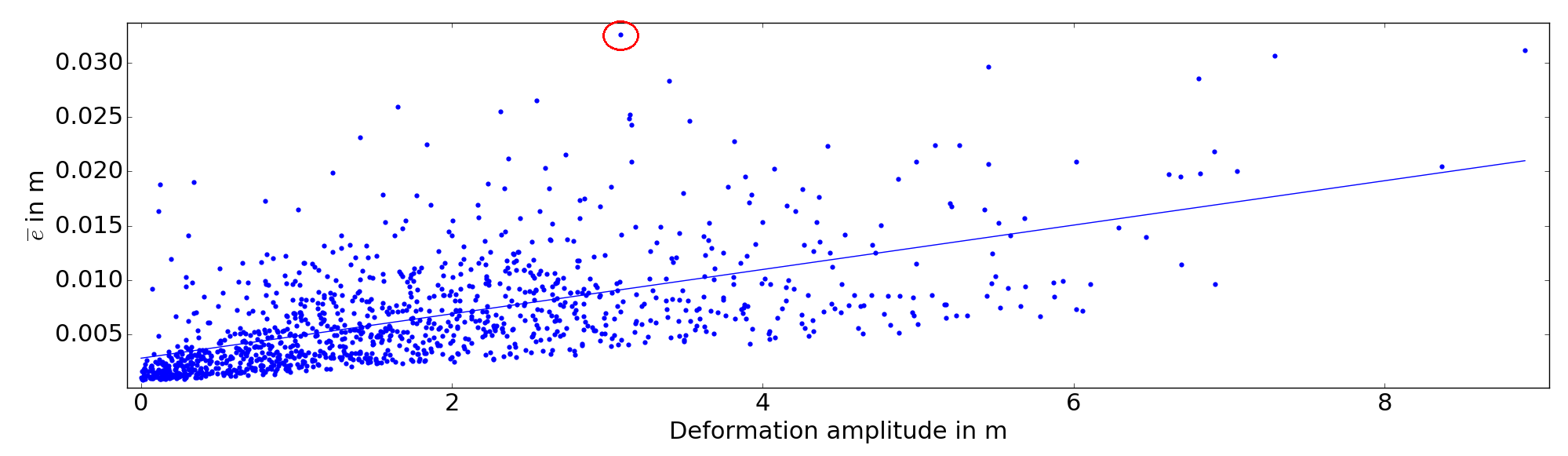}
                \caption{Sensitivity of $e$ to the deformation amplitude for the L-shape. The point cloud represents the $e$ all the samples of the testing data set. The regression line of equation $y=0.002046\times x$ shows the low sensitivity of the U-Mesh to the deformation range. Maximal error sample highlighted in red and shown in Fig.~\ref{fig:outlier}.}
             \label{fig:errors_large_L}
            \end{figure}

\begin{figure}
                \centering
                \includegraphics[width=0.1\textwidth]{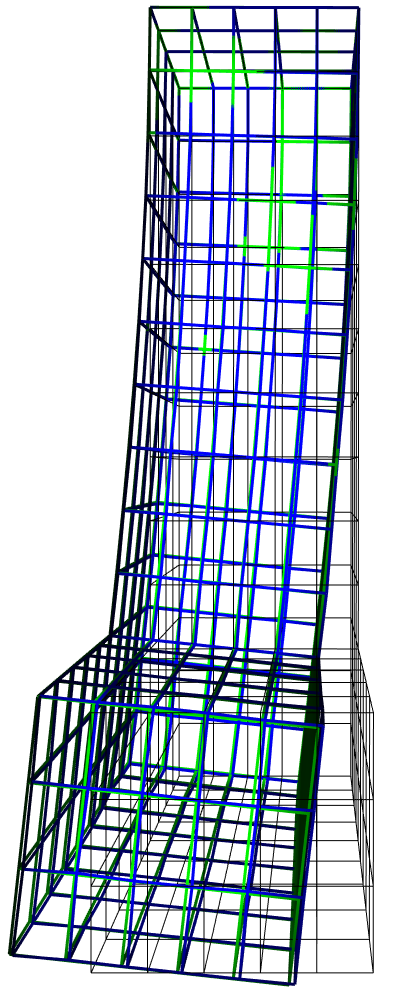}\includegraphics[width=0.3\textwidth]{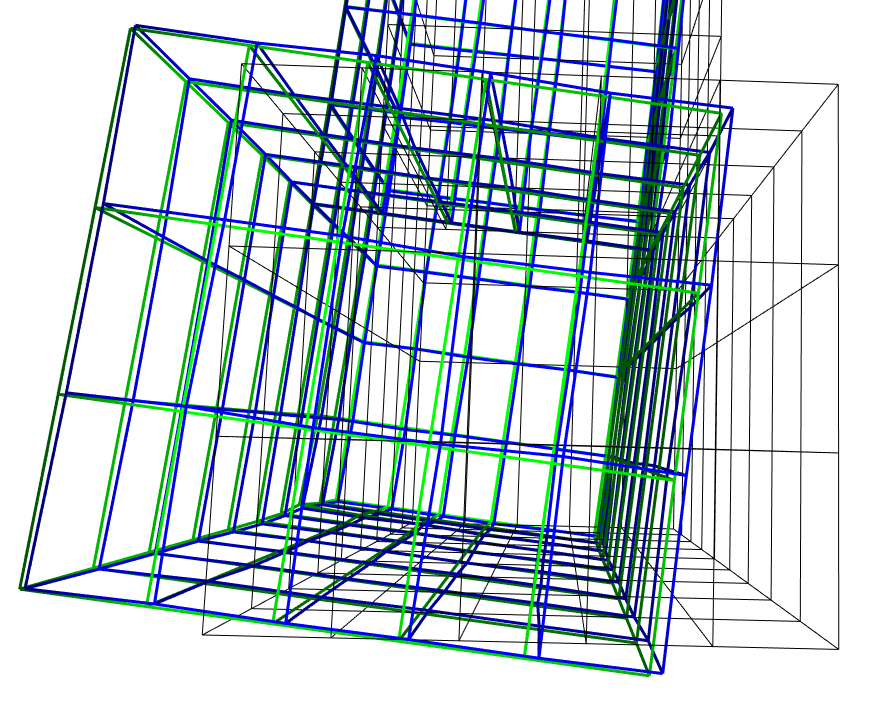}
                \caption{Maximal error sample ($e=0.0327\,m$ for a nodal deformation reaching $3.08\,m$). Reference solution is in green and U-Mesh prediction is in blue.}
                \label{fig:outlier}
            \end{figure}

We apply U-Mesh on an L-shape of size $28.424 \times 10 \times 40~m^3 $ discretized in $335$ H8 elements. Since the U-Net requires a regular grid as input, the L-shaped object is embedded in a regular grid (with zero-padding). The Young's modulus is equal to $500~Pa$ and the Poisson's ratio is $0.4$. To build the data set, external forces ranging from $0$ to $40\,N$ are applied on the bottom face of the L (with $\Lambda = 100$) up to a total of $6,000$ samples ($N=4,800$ and $M=1,200$). We train a U-Net with 128 channels in the first layer and 3 steps. In Fig.~\ref{fig:L_bounded_40} are shown some samples of deformed L-shapes. The U-Mesh output is in blue and the reference solution is in green. In order to perceive the amount of deformation, the rest position is also shown (thin grey lines). The average and maximal errors are given in Table \ref{err_large_L}, and the prediction times are in the same range as for the beam scenario. The average error is equal to $0.00648\,m$ where the maximal deformation is $8.9016\,m$. The slope of the regression in Fig.~\ref{fig:errors_large_L} shows that the increase of the error with the deformation amplitude is controlled. It is worth noting that the outliers of this graph (such as the one marked in red) still correspond to small errors (see Fig.~\ref{fig:outlier}).

             \begin{table}
            \begin{center}
              \begin{tabular}{ |c|c|c||c|c||c|c| } 
               \hline
               c & k & FSS  & $\overline{e}$ & $\sigma$($e$)&pred\_t  & train\_t \\ 
                 &  & & in m  & in m & in ms &  in min\\ 
                \hline
             128&3&1024 &0.00648&0.00493&3.2& 97\\
               \hline
              \end{tabular}
            \caption{Error measures computed over the testing data set ($M=1,200$ samples) for the L-shaped object. }
            \label{err_large_L}
            \end{center}
            \end{table}
            
\subsection{U-Mesh applied to a liver shape}
\begin{figure}
    \centering
    \subfigure[\label{fig:sg}]{\includegraphics[width=0.25\textwidth]{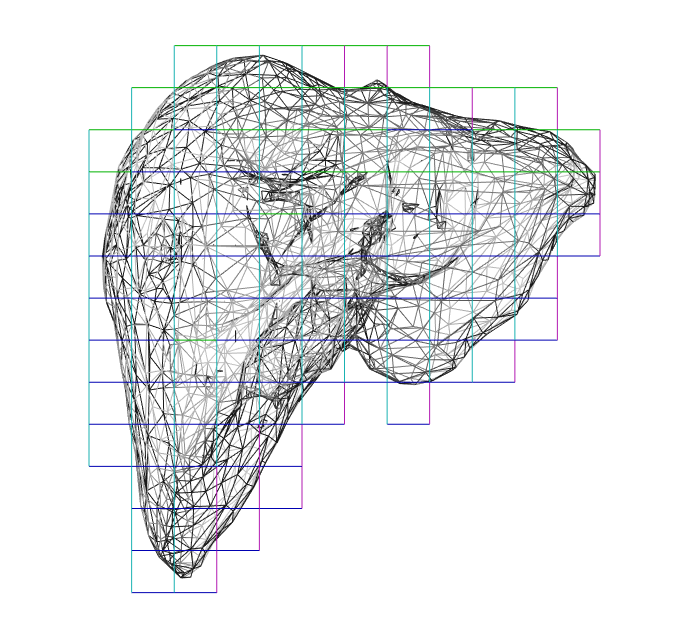}}
    \subfigure[\label{fig:rg}]{\includegraphics[width=0.2\textwidth]{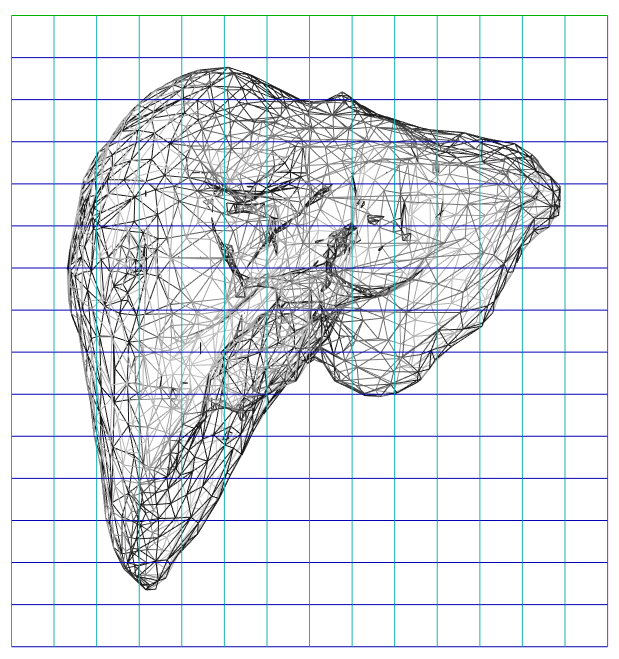}}
    \caption{(a) Hexahedral simulation sparse grid of 1109 nodes. (b) Input to the U-Net: regular grid of size $16 \times 15 \times 16 $. }
\end{figure}

To demonstrate that our method can be applied to any kind of geometry and topology, we selected a liver model to showcase the potential of U-Mesh in the field of computer-aided surgery. A surface mesh is obtained from a pre-operative CT scan of a human liver. As the method currently requires a regular grid as input,  we propose to embed the surface mesh into a sparse hexahedral grid (Fig.~\ref{fig:sg}), that is in turn embedded into a regular grid (Fig.~\ref{fig:rg}).  This sparse grid consists of rectangular cuboid cells. Cells that are overlapping the domain boundary are kept, therefore approximating the exact shape and volume of the object. Obviously, the smaller the grid, the smaller is the difference between the exact volume and the one represented by the sparse grid. If needed, it is possible to correctly account for the mesh boundary by using a more advanced integration method, as in \citep{paulus} for instance. This is however not within the scope of our paper. 

Dirichlet boundary conditions are then added by fixing 54 nodes in the area separating the two lobes to mimic the effect of the vascular tree and of the falciform ligament of the liver \citep{liver}. Normal forces of random magnitudes are computed on the liver surface and applied on the hexahedral grid through a mapping. 
Only one force is applied at each time step on a small region of the surface. We decided to limit the size of the data set to fit the time requirements of a clinical routine where sometimes only a few hours are available between the pre-operative CT scans and the surgery. Hence a data set of only $2,000$ samples is generated in $135\,min$. $N=1,600$ samples are used to train the network in $149\,min$ and $M=400$ samples are left for validation. 

\begin{figure}[h]
                \centering
                \includegraphics[width=0.5\textwidth]{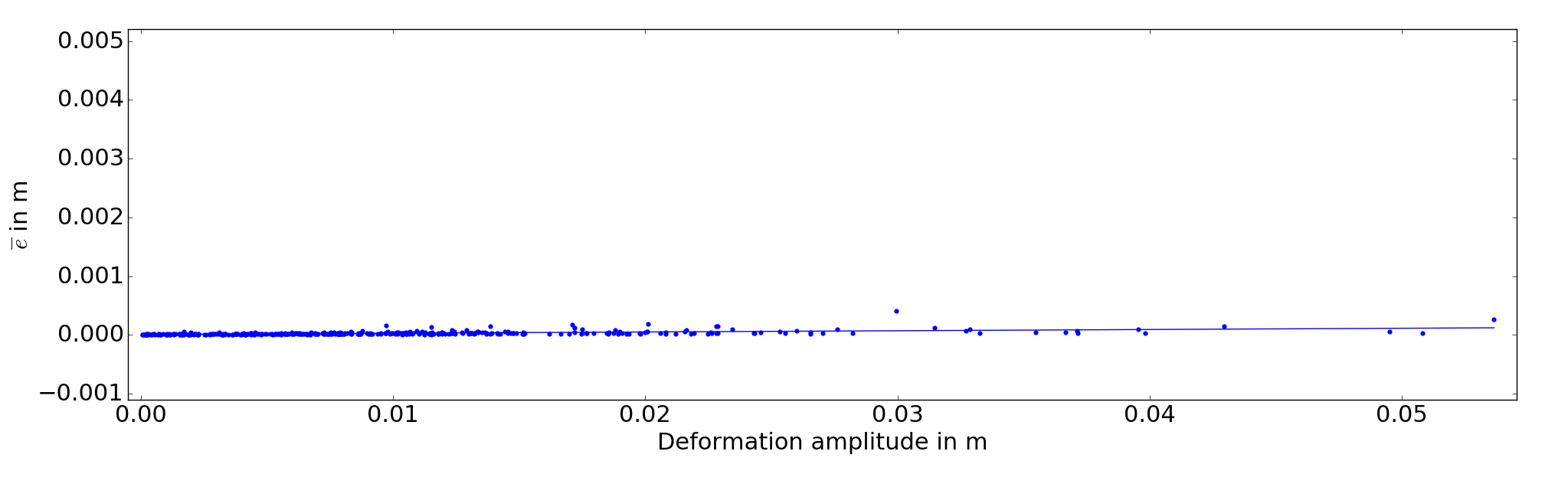}
                \caption{Sensitivity of $e$ to the deformation amplitude for the liver. The point cloud represents the error $e$ of all the samples of the testing data set. The regression line of equation $y=0.0021\times x$ shows the low sensitivity of the U-Mesh to the deformation range.}
             \label{fig:line_liver}
            \end{figure}

The metrics obtained on the validation set are reported in Table \ref{tab:liver}. The length of the liver is $0.2\,m$. The Young's modulus $Y$ is set to $5,000\,Pa$ and the Poisson's ratio to $0.48$. Our hexahedral grid has 1109 nodes forming 732 H8 elements. The maximal error is of only $4.07e-04\,m$ for a maximal deformation of $0.0536\,m$. The outputs are predicted in only $3\,ms$. In Fig.~\ref{fig:liver} are shown some samples of U-Mesh-deformed livers and their corresponding relative errors computed at one of the lobe tips. The output of U-Mesh is in green whereas the reference solution is in red. Furthermore, the slope of the regression in Fig.~\ref{fig:line_liver} shows that the increase of the error with the deformation amplitude is also controlled for this scenario.

        \begin{table}
            \begin{center}
              \begin{tabular}{ |c|c|c||c|c||c|c| } 
               \hline
               c & k & FSS  & $\overline{e}$ & $\sigma$($e$)&pred\_t  & train\_t \\ 
                 &  & & in m  & in m & in ms &  in min\\ 
                \hline
             128&3&1024 & 2.89e-05 & 3.41e-05 & 3 & 149 \\
               \hline
              \end{tabular}
            \caption{Error measures on a liver of length $0.2\,m$ immersed in a $16 \times 15 \times 16 $ grid. The maximal error is $4.07e-04\,m$ and the maximal deformation of $0.0536\,m$. }
            \label{tab:liver}
            \end{center}
            \end{table}
            
\begin{figure}[h]
    \centering
     \subfigure[Relative $l^2$ norm  3.2\%\label{fig:27}]{\includegraphics[height=33mm]{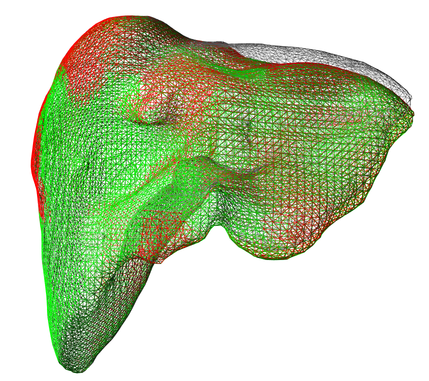}}
    \subfigure[Relative $l^2$ norm  2.6\%\label{fig:36}]{\includegraphics[height=33mm]{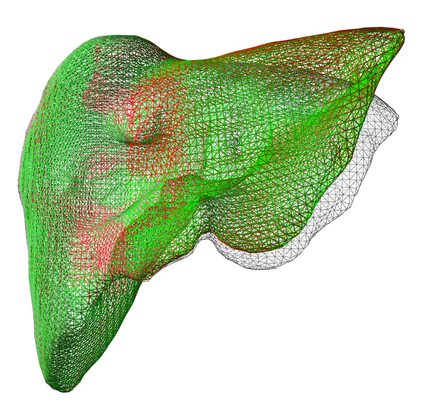}}
\subfigure[Relative $l^2$ norm  8.5\%\label{fig:86}]{\includegraphics[height=33mm]{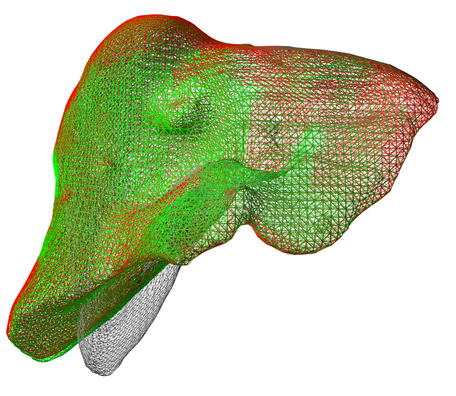}}
    \subfigure[Relative $l^2$ norm  2.4\%\label{fig:177}]{\includegraphics[height=33mm]{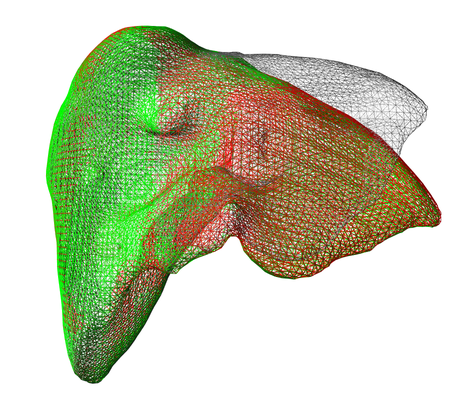}}
    \caption{Various liver samples from the testing data set and corresponding relative errors computed on the tip of the deformed lobe. The rest shape of the liver is shown in grey.}
    \label{fig:liver}
 \end{figure}   
 
 \begin{figure}[h]
    \centering
    \includegraphics[width=0.5\textwidth]{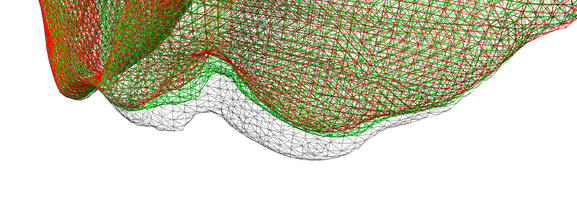}
    \caption{Sample with maximal maximal nodal error ($0.0153\, m$) for the hexahedral mesh. The reference solution is shown in red and the U-Mesh predictions is in green. The rest shape is shown in grey.}
    \label{fig:liver}
 \end{figure} 
 
\begin{figure}[h]
    \centering
    \subfigure[Relative $l^2$ norm  2.4\%\label{fig:27}]{\includegraphics[height=33mm]{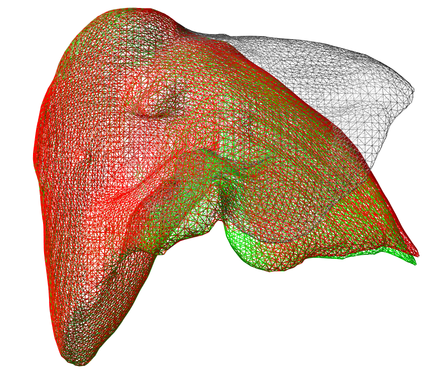}}
    \subfigure[Relative $l^2$ norm  1.9\%\label{fig:36}]{\includegraphics[height=33mm]{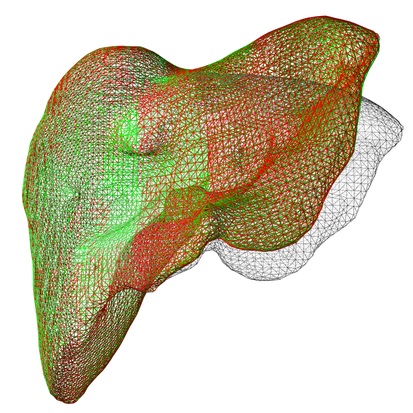}}
\subfigure[Relative $l^2$ norm  8.4\%\label{fig:86}]{\includegraphics[height=33mm]{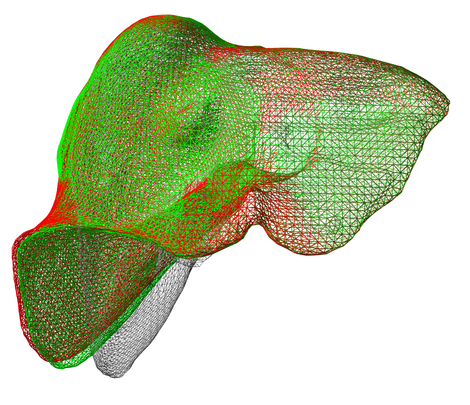}}
    \subfigure[Relative $l^2$ norm  10\%\label{fig:90}]{\includegraphics[height=33mm]{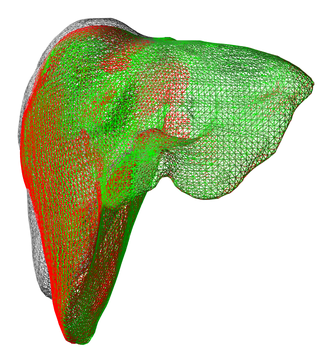}}
   
    \caption{Various liver samples from the testing data set and corresponding relative errors computed on the tip of the deformed lobe for the tetrahedral topology. The rest shape of the liver is shown in grey.}
    \label{fig:tetra_results}
 \end{figure}

\begin{figure}[h]
    \centering
    \subfigure[Maximal average nodal error.\label{fig:227}]{\includegraphics[width=0.2\textwidth]{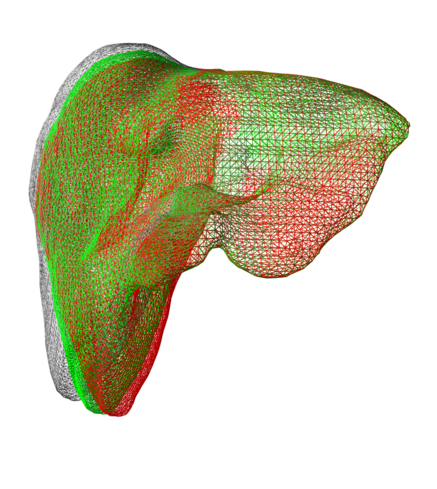}}
    \subfigure[Regular grid mapped onto a FE tetrahedral mesh. \label{fig:tetra}]{\includegraphics[width=0.22\textwidth]{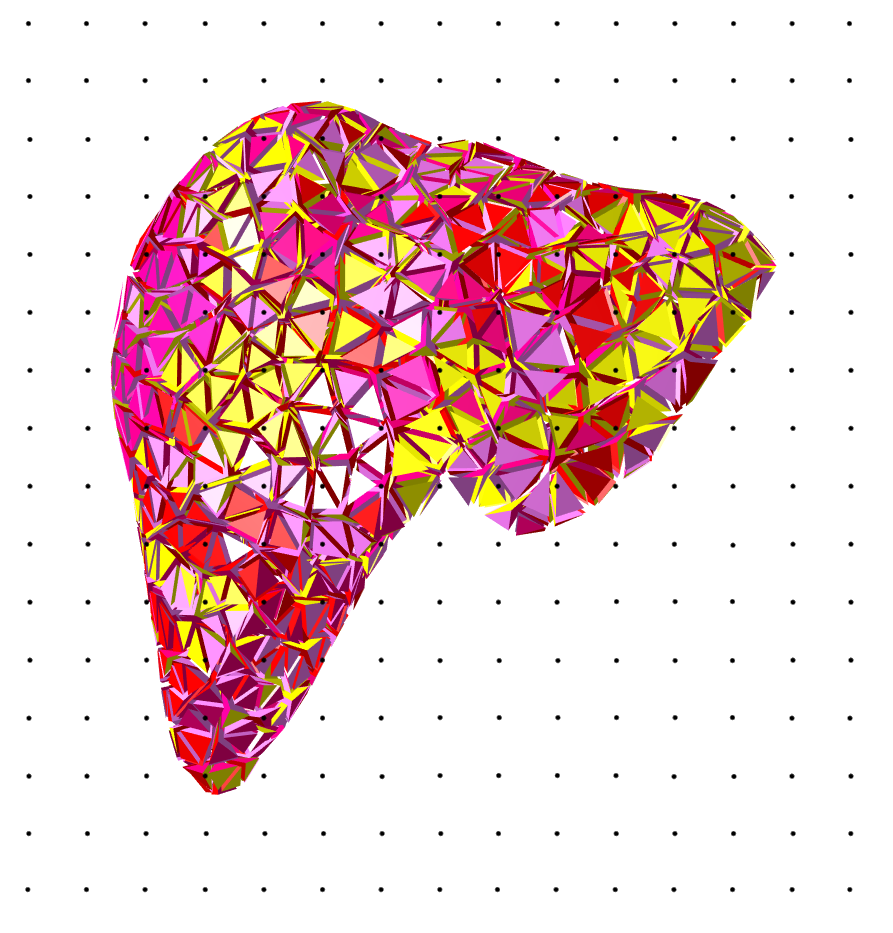}}

    \caption{Maximal error for liver discretized in tetrahedral elements.}
    \label{fig:max_error_liver_tetra}
 \end{figure}

The input to the network needs to have a grid-like structure and this might be seen as a limitation of our approach. However, through this example, we will demonstrate that it can work with any kind of FE mesh. In this part the liver geometry is discretized into 4859 tetrahedral elements (1059 nodes in total). A $16  \times 15 \times 16$ regular grid is mapped onto the tetrahedral mesh and follows the deformation of the FEM mesh (see Fig.~\ref{fig:tetra}). Only the nodes of the grid that are inside the liver volume are mapped. The outer nodes are zero-valued. Similarly to previous scenarios, normal forces of random magnitudes are applied to the surface of the liver and mapped to the regular grid in order to generate a data set of $2,000$ samples. The results reported on Table \ref{tab:tetra_liver} and on Fig.~\ref{fig:tetra_results} and \ref{fig:max_error_liver_tetra} demonstrate the ability of our method to predict deformations for any kind of topology.


        \begin{table}
            \begin{center}
              \begin{tabular}{ |c|c|c||c|c||c|c| } 
               \hline
               c & k & FSS  & $\overline{e}$ & $\sigma$($e$)&pred\_t  & train\_t \\ 
                 &  & & in m  & in m & in ms &  in min\\ 
                \hline
             128&3&1024 & 5.33e-05 & 6.03e-05 & 3 & 149 \\
               \hline
              \end{tabular}
            \caption{Error measures on a liver of length $0.2\,m$ discretized with tetrahedral elements. The maximal error in the testing data set is equal to $4.9e-04\,m$ (sample in Fig.~\ref{fig:227}) for a maximal deformation over the testing data set of $0.088\,m$. }
            \label{tab:tetra_liver}
            \end{center}
            \end{table}
            
The obtained results highlight the potential of our method in applications where high accuracy and computational efficiency are demanded. Overall, we have seen that U-Mesh predicts deformations for different geometries and mesh resolutions, with small controlled errors, in a very short amount of time (about $3~ms$).

\section{ Discussion}
\label{sec:discussion}

\subsection{Comparison of U-Mesh and POD}
        
In this section we compare the predictions made by U-Mesh to the simulations computed on a reduced model using POD. We used the POD code available at \url{https://github.com/SofaDefrost/ModelOrderReduction} that works as a plugin of the SOFA framework \citep{sofa}. The POD consists in three phases. First, an offline phase where all the potential movements of the beam are sampled and stored in the so-called snapshot. This offline phase is the equivalent to the data generation phase in U-Mesh and is also computationally intensive since it performs many fine simulations. In a second phase, the snapshot space is condensed in a reduced basis using singular value decomposition and a truncation \citep{Goury2018}. This phase "corresponds" to the training of the U-Net and is generally faster. Finally, the resulting reduced model allows for faster simulations since there are fewer degrees of freedom. We applied the simple POD and the hyperreduced-POD (HPOD) to the fine beam depicted previously (see Fig.~\ref{fig:fine}) in order to compare the performance of POD and U-Mesh. 

We first compare the computation times for a given accuracy. The truncature error of the POD was set such that the mean norm error obtained with POD is similar to the one obtained with U-Mesh. To reach the desired accuracy in the considered deformation range, 3 modes were preserved. Computation times are reported in Table \ref{umesh_vs_pod_cpu}. With the selected number of modes, POD is about 6 times faster than the full FEM model whereas U-Mesh is more than 200 times faster than the full FEM model.


Let us now compare the relative errors of the two methods for a given computation time. The fastest version of the reduced model is the one considering only one deformation mode and using hyperreduction. As presented in Table \ref{umesh_vs_pod_error}, HPOD can compute deformations in $5\,ms$ but with an error that is 14 times larger than the one produced by U-Mesh.

     \begin{table}
     \begin{center}
    \begin{tabular}{|c|c|c|}
    \hline
          & Prediction time & $e$  \\
          & in s &  in m \\
    \hline
         FEM & 0.740 & 0.000 \\
         POD & 0.120 & 0.006\\
         U-Mesh & 0.003 & 0.006\\
    \hline
    \end{tabular}
     \caption{Comparison at same error: computation times of the full FEM model, of the POD and of the U-Mesh, for the same mean norm error $e$. The number of modes used in the POD is 3.}
    \label{umesh_vs_pod_cpu}
    \end{center}
    \end{table}

    \begin{table}
     \begin{center}
    \begin{tabular}{|c|c|c|}
    \hline
          & Prediction time & $e$ \\
& in s &  in m \\
    \hline
         FEM & 0.740 & 0.000 \\
         HPOD & 0.005 & 0.084\\
         U-Mesh & 0.003 & 0.006\\
    \hline
    \end{tabular}
 \caption{Comparison at similar computation time: the mean norm errors of the full FEM model, of the HPOD and of the U-Mesh are given, for a computation time in the range of the millisecond. Only one mode is kept in the HPOD.}
    \label{umesh_vs_pod_error}
    \end{center}
    \end{table}
    
    \subsection{Extension to surgical scenarios}
    
    We are planning to extend our method to surgical simulation and in particular to guidance during hepatic surgery where our approach can provide a mean to register the preoperative model in real-time. Before surgery, a biomechanical model of the organ can be built based on the anatomical geometry segmented from a CT-scan of the patient. The training samples would be generated and used to train the network in an offline phase. Assuming we know the location and the force applied by the surgical tool, this force could be mapped to the U-Mesh grid and given to the network so that it predicts the resulting deformation in real-time. However, in most surgical scenarios, the force applied by the instrument is unknown and the only available information is a partial surface deformation (that could be acquired by different imaging techniques). In this configuration, the network would learn the relationship between surface displacements and volumetric displacements. If, besides the displacement field, we also want to estimate the stress in the organ, we could use more complex hyperelastic laws or expand our model to account for viscoelasticity. Handling models such as Ogden or Mooney-Rivlin would require no change in our method. However to handle viscoelasticity, we would need to include a time term and the current state of the system in the network input as done by \cite{Meister}. 
     
    Another scenario worth investigating is the case of contacts between anatomical structures. Interaction between objects can be seen as external forces applied to their surfaces (the alternative option being to solve interactions through position constraints). Assuming we have two objects embedded in two U-Mesh grids, we can compute their motion until a contact is detected and then apply a simple penalty-based contact response. This contact response is a force applied on the surface of each object to cancel out their interpenetration. Using this force, we could then compute the deformation of each object following our method.

    \subsection{Current limitations}
 
    Despite the promising results of our method, there are some worth mentioning limitations. With the current U-Mesh, it is not possible to make very accurate predictions when applying a force somewhere out of the sampled input domain of the training data set. This limitation is also inherent to POD. In the same manner, we are restricted to the geometry used to train the network. 
    
    Another limitation of the method is the expensive offline phase. The data generation can be extremely time consuming in particular when considering large meshes or when more complex input sequences are needed. It goes without saying that the larger the data sets, the longer the training. Hence it is important to build smart data generation strategies to cover all the force ranges without being exhaustive (to reduce data generation and training times). An option would be to perform intelligent sampling for the data set generation, such as Latin-Hypercube sampling. Moreover, we performed a little study on the sensitivity of the method to the Young's modulus variations in the training data set. We noticed that U-Mesh is more accurate for stiffer objects. This is an important result to keep in mind for the future \textit{smart data generation.} Indeed, in a low deformation regime less data is needed to reach good accuracy.

\section{Conclusion}
In this paper we have proposed a U-Net architecture that can learn the relation function between an input force and an output deformation for various geometries and make predictions with high accuracy in a record amount of time. The take-home-message of this work is that for a given network architecture, the prediction time is nearly constant (and very short), irregardless of the size of the problem. Furthermore the accuracy of the prediction, which depends on the quality and the size of the data set, is controllable since we generate this data.
We believe that such an approach has a tremendous potential for problems requiring very fast simulations of objects undergoing interactions. Such interactions could be user-driven or the result of contacts with other structures. 

Yet, there are several directions to investigate to make this approach more broadly usable, in particular in the context of biomechanics, where material parameters, boundary conditions and geometries can vary from one patient to another. To this end, we will be investigating transfer learning as a means to refine a neural network pre-trained on an average model. 
\vspace{-0.25cm}
\section{Aknowledgments}
The authors would like to thank Jean-Nicolas Brunet for proofreading the manuscript. The authors also report no conflict of interest.

\noindent{Conflicts of interest:}
None




\end{document}